\documentclass{article}

\usepackage[sglblindworkshop, final]{ai4nextg_neurips_2025}

\usepackage[utf8]{inputenc} 
\usepackage[T1]{fontenc}    
\usepackage{hyperref}       
\usepackage{url}            
\usepackage{booktabs}       
\usepackage{amsfonts}       
\usepackage{nicefrac}       
\usepackage{microtype}      
\usepackage{xcolor}         

\usepackage{amsmath,amssymb}
\usepackage{graphicx}
\usepackage{subfigure}
\usepackage[most]{tcolorbox} 
\usepackage{natbib}

\title{Constrained Network Slice Assignment via Large Language Models}
\author{
  Sagar Sudhakara \\
  University of Southern California \\
  \texttt{sagarsud@usc.edu}
  \And
  Pankaj Rajak\\
  University of Southern California\\
  \texttt{rajak@usc.edu}
}
\begin{document}
\maketitle

\begin{abstract}
Modern networks support \emph{network slicing}, which partitions physical infrastructure into virtual slices tailored to different service requirements (e.g., high bandwidth, low latency). Optimally allocating users to slices is a constrained optimization problem that traditionally requires complex algorithms. In this paper, we explore the use of Large Language Models (LLMs) to tackle the radio resource allocation for network slicing. We focus on two approaches: (1) using an LLM in a zero-shot setting to directly assign user service requests to slices, and (2) formulating an integer programming model where the LLM provides semantic insight by estimating similarity between requests. Our experiments show that an LLM, even with zero-shot prompting, can produce a reasonable first draft of slice assignments, though it may violate some capacity or latency constraints. We then incorporate the LLM’s understanding of service requirements into an optimization solver to generate an improved allocation. The results demonstrate that LLM-guided grouping of requests (based on minimal textual input) achieves performance comparable to traditional methods using detailed numerical data in terms of resource utilization and slice isolation. While the LLM alone is not perfect in satisfying all constraints, it significantly reduces the search space and, in conjunction with exact solvers, highlights a promising collaboration for efficient 5G network slicing resource allocation.
\end{abstract}

\section{Introduction}
Network slicing empowers mobile networks to accommodate diverse application requirements by creating isolated virtual networks (slices) on shared infrastructure. For example, an enhanced mobile broadband (eMBB) slice can support high-throughput services like streaming video, while an ultra-reliable low-latency communications (URLLC) slice caters to mission-critical services needing minimal delay. A third category, massive Machine-Type Communications (mMTC), serves large numbers of IoT devices with sporadic low-data transmissions. Slicing enables differentiated Quality of Service (QoS) guarantees for each category. However, effectively managing and allocating resources to these slices under strict capacity and latency constraints is challenging. The allocation must ensure service isolation (preventing slices from negatively impacting each other’s performance) while optimizing overall resource usage. This often leads to a complex constrained optimization problem involving many parameters and requirements.

Traditionally, determining how to map user demands to slices (and how to distribute radio resources among slices) relies on solving large-scale optimization formulations or using heuristic algorithms. Such constrained optimization tasks are computationally intensive and require careful tuning of algorithms for the unique conditions of the network. Even integer linear programming (ILP) solutions struggle to scale beyond small problem instances due to the explosion of decision variables and constraints in realistic networks. Network engineers typically perform slice resource allocation manually or with custom tools, especially during initial slice configuration or when adjusting slices for new services.

Recently, Large Language Models have shown remarkable success in various domains requiring reasoning and planning. LLMs can parse complex instructions and even perform certain logical or arithmetic tasks by virtue of their training on vast amounts of data. This raises a natural question: \textit{Can an LLM help solve the network slicing resource allocation problem?} Specifically, we investigate whether an LLM (such as Claude models) can propose feasible assignments of users to slices under given constraints, and how its understanding of request semantics might be leveraged in optimization. Our investigation revolves around two primary approaches, inspired by prior work in other scheduling domains: (1) using zero-shot prompting of the LLM to generate a slice assignment directly, and (2) using the LLM’s ability to evaluate similarity or compatibility between requests as input to an integer programming (IP) solver for optimal allocation.

By applying LLMs to this wireless domain, we aim to assess their capacity to handle the multi-criteria decision-making inherent in 5G slicing. In the following, we formalize the slicing allocation problem and then describe experiments with both the zero-shot LLM approach and the IP optimization approach. We evaluate results based on resource utilization, constraint satisfaction, and slice isolation, and discuss how a combined LLM+solver strategy can harness the strengths of AI and traditional optimization for future network management.

\section{Related Work} 
\paragraph{Network Slicing Optimization.} 5G network slicing has traditionally been tackled with rigorous optimization and machine learning methods. Researchers have formulated slice resource allocation as constrained integer programming problems (ILP/MILP), ensuring optimal or near-optimal fulfillment of slice capacity and QoS constraints. For example, classical approaches use ILP to map service requests to slices under capacity and latency limits, or to embed end-to-end slices in substrate networks \citep{Li2020}. These methods guarantee feasible allocations with provable optimality under simplified conditions, but they face scalability issues as problem size grows. To mitigate complexity, heuristic and learning-based algorithms have been proposed. Recent works combine model-driven slice modeling with gradient-based optimization (e.g., the MicroOpt framework for dynamic slice resource tuning) to improve efficiency \citep{Sulaiman2025MicroOpt}. Deep learning and reinforcement learning have also been applied to slicing—\citet{Bega2019} use deep neural policies for admission control in sliced networks—trading optimality for adaptability. Overall, existing solutions require extensive numerical parameters or training, motivating new paradigms for more scalable and semantic-aware resource allocation. \paragraph{LLMs and Foundation Models in Networks.} The rise of large language models (LLMs) has prompted exploration of foundation models for network management and wireless systems. Very recent studies demonstrate that prompt-guided LLMs can serve as generalist problem-solvers in networking tasks without task-specific training. \citet{Zhou2025WirelessLLM} provide an overview of LLM-enabled wireless networks, showing that carefully crafted prompts (e.g. iterative or self-refining prompts) allow GPT-style models to perform tasks like base station configuration and network performance prediction. In the networking community, \citet{Wu2024NetLLM} propose \emph{NetLLM}, a framework adapting pre-trained LLMs to networking problems by incorporating domain-specific input encoders; they report that a single foundation model can achieve competitive results on diverse tasks (congestion control, video streaming, job scheduling) compared to specialized deep models. For wireless RAN control, \citet{Wu2025LLMxApp} present an LLM-powered xApp in the O-RAN architecture that translates high-level service intents into real-time radio resource allocation decisions. Their system, which prompts an LLM agent with slice QoS requirements, improves slice throughput and reliability, illustrating the promise of LLM-driven control in 5G networks. To our knowledge, however, the use of LLMs for the specific problem of 5G slice assignment has not been explored prior to our work. \paragraph{LLMs for Structured Decision-Making.} Beyond networking, there is growing interest in using LLMs for structured or constrained decision problems, bridging generative AI with combinatorial optimization. Recent works have examined whether LLMs can act as planners or solvers when given a problem description in natural language. For example, \citet{Abgaryan2024} fine-tune a language model to produce solutions to job-shop scheduling problems and show that the LLM can generate valid schedules with makespans competitive to heuristic methods. Other studies incorporate reasoning techniques (chain-of-thought prompting, self-consistency) to guide LLMs in solving puzzles and planning tasks under constraints \citep{Huang2024}. These efforts hint at a broader trend: LLMs can serve as high-level decision-makers by leveraging their world knowledge and reasoning, while traditional optimization or search algorithms ensure constraint satisfaction. Our approach follows this hybrid philosophy. We use an LLM to quickly draft a slice allocation based on semantic understanding of service descriptions, and then refine it with an exact solver to enforce hard capacity and latency constraints. This combination positions our work at the intersection of foundation models and wireless network control, where LLMs provide intelligent heuristics and insights that complement conventional optimization. In summary, we extend prior research on generative AI for decision-making into the domain of 5G resource allocation, demonstrating that even a zero-shot LLM can contribute meaningfully to complex network optimization when paired with constraint-solvers.

\section{Methodology}
\subsection{Problem Formulation}
We focus on automating the allocation of user service requests to predetermined network slices, treating it as a constrained grouping and assignment problem. The network provides a fixed set of slice types (for instance, slices optimized for eMBB, URLLC, and mMTC services), each with certain resource capacities (e.g., bandwidth or resource blocks) and QoS attributes (such as a maximum allowable latency). Each incoming user request (or service flow) has an associated resource demand (amount of radio resource units needed) and a QoS requirement (e.g., a maximum latency it can tolerate). The goal is to assign every request to one of the available slices such that all constraints are respected. Key constraints include:
\begin{itemize}
    \item \textbf{Slice capacity and Latency requirement:} The total demand of requests allocated to a slice must not exceed that slice’s capacity (ensuring no slice is over-committed). A request can only be placed on a slice if that slice can meet the request’s latency requirement. For example, a URLLC-type request requiring ultra-low latency can only be assigned to a slice with sufficiently low latency (and not to a higher-latency eMBB slice).
    \item \textbf{Slice availability:} The number and type of slices are fixed in advance (determined by the operator’s provisioning). No new slices may be created on the fly, and each request must fit into one of the existing slices.
    \item \textbf{Complete assignment:} All user requests must be assigned to some slice (we do not allow rejecting or dropping a request in this scenario).
\end{itemize}

Formally, we have a set of slices $M$ (each with capacity $C_m$ and latency attribute $\ell^{(m)}$) and a set of requests $N$ (each with demand $d_i$ and required latency $\ell_i$). The assignment can be represented by binary decision variables $x_{i,m}$ indicating if request $i$ is placed in slice $m$. The optimization objective can vary – one could aim to maximize overall resource utilization, balance load, or group “compatible” requests together. In our formulation, we consider an objective that promotes grouping similar service requests in the same slice (to exploit slice specialization), mirroring scheduling problems. This leads to an objective that constrains assigning requests of similar nature to the same slice (defined more formally below and in appendix).

This problem setup bears resemblance to constrained clustering: we are effectively clustering the $N$ requests into at most $|M|$ groups (slices), under capacity and latency constraints. Importantly, slices need to maintain isolation despite sharing the physical network; this means careful enforcement of resource limits per slice to preserve performance guarantees. We do not consider time dynamics in this work – i.e., we address a one-shot allocation of a static batch of requests to slices, analogous to creating a slicing plan for current demands. Extending to dynamic or time-varying scenarios is left for future work.

\subsection{LLM-Based Zero-Shot Assignment}
Our first approach directly prompts an LLM to create a slice assignment from scratch, given a description of the problem and the list of available slices and requests. We use Claude model as the representative LLM. The prompt (see Appendix~\ref{app:prompt} for the full template) describes the slices (including their capacities and latency characteristics) and enumerates the pending user requests with their demand and latency requirements. The LLM is asked to output an assignment of each request to a slice in a specific structured format (CSV-like with each line indicating a slice–request pairing and the allocated units).

We carefully engineered the prompt to emphasize the constraints and correct format. For instance, we explicitly list constraints (1)--(4) from above and provide an example output. We set the temperature of the LLM to 0.8 to induce variability in the outputs (simulating different possible allocation decisions). After prompt tuning, Claude was able to generate a slice assignment that obeyed the format and mostly followed the rules.

This zero-shot approach leverages the LLM’s ability to reason about each request’s description and the slice properties. Intuitively, we expect the LLM to place high-bandwidth requests in the eMBB slice, low-latency requests in the URLLC slice, etc., based on the semantic cues in the prompt. However, since the LLM is not an exact numerical solver, it may produce assignments that violate constraints (e.g., exceeding a slice’s capacity or assigning a request to an inappropriate slice). We treat such outputs as initial “draft” allocations that may need refinement. In practice, if the LLM output violated a hard constraint, we recorded the violation (and could later adjust the assignment manually or via an algorithm). The strength of this approach is that it requires no specialized training or optimization algorithm—just the general-purpose reasoning of the LLM applied to our scenario.

\subsection{LLM-Guided Integer Programming}
Our second approach is a hybrid method that leverages the strengths of LLMs and traditional solvers. The idea is to use the LLM’s deep semantic understanding of service requirements to inform an integer programming model, rather than having the LLM construct the full solution alone. This approach acknowledges that LLMs excel at semantic reasoning (e.g., understanding which services are “similar” or should be grouped) but ILP (Integer Linear Programming) excels at exact constraint enforcement and optimization.

Concretely, we incorporate a notion of \emph{request similarity} into the optimization. We define a similarity score $\text{sim}(i,j)$ between any two requests $i$ and $j$ that reflects how compatible they are to be served by the same slice. For instance, two high-bandwidth video streams (eMBB-type requests) would have a high similarity, whereas a latency-critical control signal and a high-bandwidth video might have low similarity (since one is URLLC and the other eMBB). Determining these similarity values automatically is a challenge because it requires understanding the nature of the services – something LLMs are well-equipped for. We prompt Claude model to analyze the descriptions or attributes of pairs of requests and output a qualitative judgment (which we map to a numerical score). For example, the LLM might conclude: “Request 7 and Request 12 both involve streaming video – they are similar (score 1). Request 7 and Request 15 have very different requirements – they are not similar (score 0).” By querying the LLM for all request pairs, we build an $N\times N$ similarity matrix where each entry is 1 if the LLM deems the two requests compatible (or similar enough to benefit from being in the same slice), and 0 if not.\footnote{In principle the similarity scores could be continuous values between 0 and 1 for varying degrees of similarity. In our experiments we used binary judgments for simplicity.}

With this LLM-derived similarity matrix as input, we then formulate an ILP for optimal slice allocation. The ILP (detailed in Appendix~\ref{app:IP}) introduces the same assignment variables $x_{i,m}$ as before, and additional binary variables $z_{i,j,m}$ to link assignments of request pairs (indicating if $i$ and $j$ are both assigned to slice $m$). The objective function is set to maximize the total similarity of requests that are placed in the same slice. Intuitively, this encourages the solver to group like with like (i.e., assign requests that the LLM marked as similar into the same slice), thereby aligning with the intended specialization of slices. This objective is subject to all the earlier hard constraints: each request must be assigned exactly once, no slice capacity is exceeded, and latency requirements are honored.

We solve this ILP using a standard solver (GLPK in our case). The solver finds an assignment of requests to slices that maximizes the similarity-based objective while satisfying all constraints. 

The output of the IP solver is an optimal slice allocation (if a feasible solution exists). Notably, by the construction of the ILP, any solution it finds will satisfy the capacity and latency constraints by design. Thus, the ILP solution (whether using LLM-based or baseline similarities) yields a valid allocation with almost zero constraint violations.

\section{Experiments}
\subsection{Synthetic Dataset and Experimental Setup}
Performing controlled experiments on real 5G network data can be challenging due to privacy, security, and proprietary restrictions on telecom datasets. Therefore, we constructed a synthetic dataset to evaluate our methods, ensuring it captures realistic characteristics of network slice demands while avoiding any sensitive information. By using synthetic data, we have full flexibility and avoid concerns of exposing confidential network configurations.

Our test scenario consists of three slice types (Slice~A, Slice~B, Slice~C), corresponding to eMBB, URLLC, and mMTC slices respectively. Each slice type is configured with a capacity (total resource units available) and a latency guarantee representative of its category (for example, Slice~A has high capacity and a relaxed latency threshold, Slice~B has lower capacity but ultralow latency, etc.). We then generate a set of $N$ user service requests. Each request is assigned a random resource demand and a latency requirement drawn from distributions appropriate to various service types. For instance, some requests are high-bandwidth (large $d_i$) and latency-tolerant (large $\ell_i$) mimicking eMBB traffic, whereas others are small $d_i$ but require very low $\ell_i$ (typical URLLC), and some represent mMTC (very small $d_i$ and high $\ell_i$ tolerance, but numerous in quantity). We ensure that the overall demand is such that not all requests can trivially fit into one slice, making the allocation problem non-trivial (some slices will be oversubscribed if allocation is done naively).

Using this scenario, we evaluated both methods. For the LLM zero-shot approach, we input the scenario description to Claude models. The prompt used followed the template in Appendix~\ref{app:prompt}. We randomly shuffled the order of requests in the prompt for each run to prevent the LLM from relying on any incidental ordering. We ran multiple independent trials of the LLM (with temperature 0.8) to observe variability and collect statistics on its performance. For the ILP approach, we used the GLPK solver to solve the optimization described in Section 3.3. We obtained the similarity matrix from the LLM as described earlier (by querying Claude model about request similarities prior to solving) and solved the ILP to near optimality. 

\subsection{Results and Analysis}
We first examine the performance of the zero-shot LLM solution. Encouragingly,  Claude models was able to assign \emph{all} user requests to slices in every trial (completeness was essentially 100\%), indicating that it understood the instruction that no request should be left unassigned. The assignments produced were usually intuitive: high-demand, high-bandwidth requests were typically placed into the eMBB slice, and low-latency requests were placed into the URLLC slice, in line with expectations. This demonstrates the LLM’s capability to interpret the problem and make reasonable allocation decisions purely based on the described constraints and request attributes.

\begin{table}[h!]
\centering
\caption{Completeness and Homogeneity across Claude variants (mean $\pm$ std).}
\small
\begin{tabular}{lcc}
\toprule
\textbf{Model} & \textbf{Completeness (\%)} & \textbf{Homogeneity} \\
\midrule
claude-3-haiku & 100.00 ± 0.00 & 0.35 ± 0.27 \\
claude-3-sonnet & 99.50 ± 1.50 & 1.00 ± 0.00 \\
claude-3-5-sonnet & 100.00 ± 0.00 & 0.50 ± 0.50 \\
\bottomrule
\end{tabular}
\end{table}

\begin{figure}[h!]
    \centering
    \includegraphics[width=0.8\linewidth]{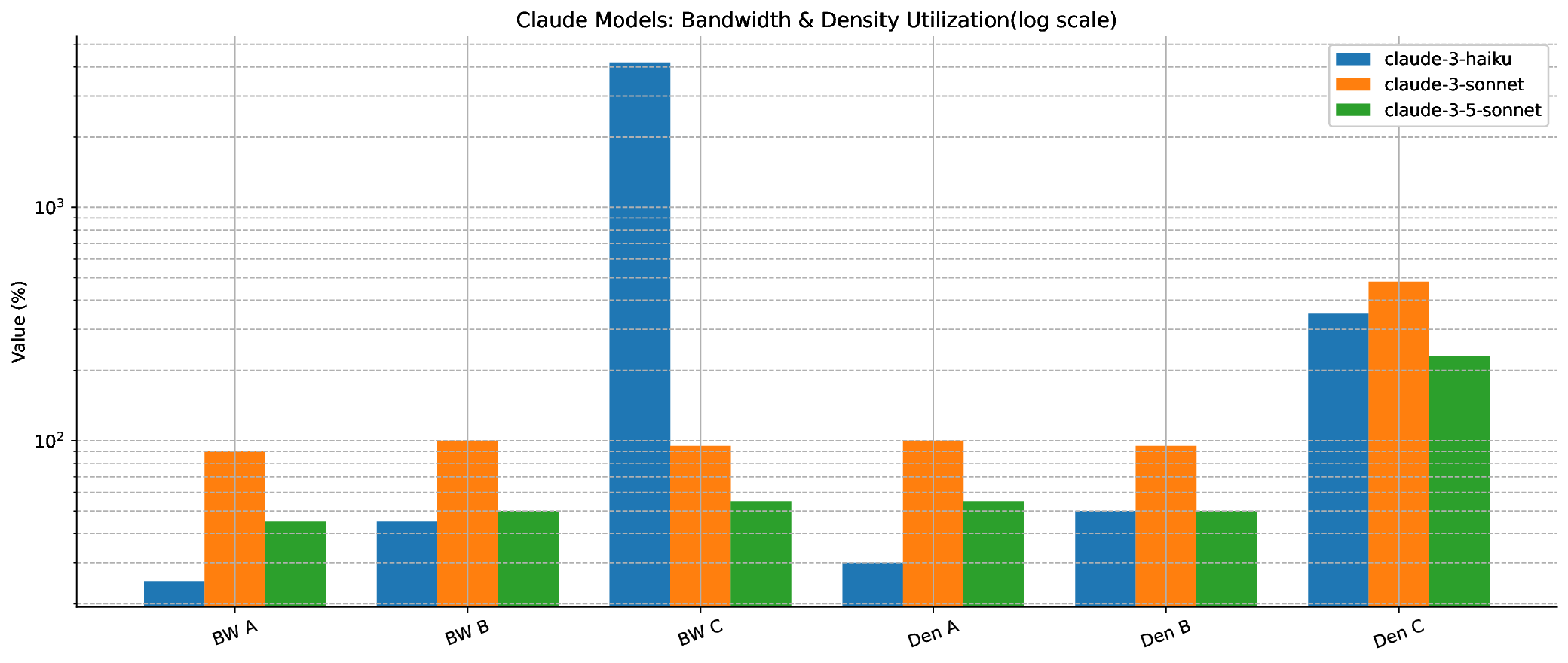}
    \caption{Bandwidth and Density Utilization for Claude variants.}
    \label{fig:claude_bw_den}
\end{figure}
Across the three Claude variants, completeness was consistently high ($\approx$100\%), while homogeneity varied widely, peaking at 1.0 for \textit{claude-3-sonnet}. The \textit{sonnet} model also showed balanced bandwidth and density utilization near 90--100\% for all slices, whereas \textit{haiku} had extreme overutilization in Slice~C and underuse in Slices~A/B, and \textit{claude-3.5-sonnet} exhibited moderate, more balanced utilization.

\begin{figure}[h!]
    \centering
    \includegraphics[width=0.9\linewidth]{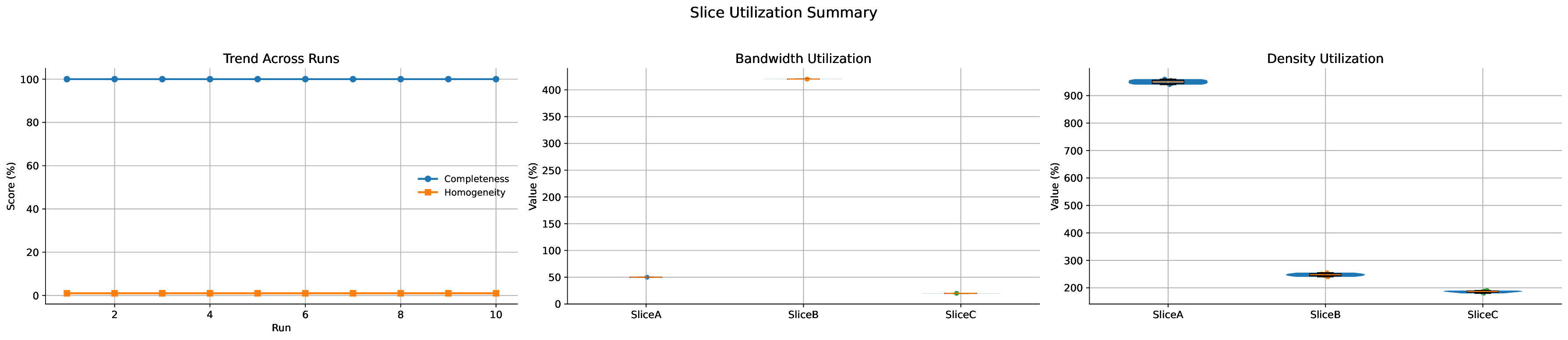}
    \caption{Performance metrics for the ILP+LLM allocation method. Left: Completeness and homogeneity scores across 10 runs. Center: Distribution of bandwidth utilization across Slice A, Slice B, and Slice C. Right: Distribution of density utilization across Slice A, Slice B, and Slice C.
}
    \label{fig:claude_bw_den}
\end{figure}

\noindent\textbf{Comparison of Methods:}  
While direct LLM-based allocation can produce high-quality semantic groupings, it may suffer from occasional constraint violations or imbalances in resource utilization.  
The ILP+LLM hybrid approach combines the LLM’s strength in semantic clustering with the ILP solver’s rigorous constraint enforcement, yielding allocations that are both feasible and near-optimal.  
This synergy ensures 100\% completeness and balanced utilization across slices, outperforming either method alone and providing a robust, scalable solution for network slicing.

\section{Conclusion}
Our investigation demonstrates that large language models (LLMs) can rapidly produce sensible, structured slice allocation plans for 5G network slicing in a zero-shot setting, even without domain-specific training. However, when operating alone, current LLMs are prone to occasional constraint violations and imbalances in resource utilization, particularly at larger problem scales.  

By integrating LLM-generated semantic groupings with an integer linear programming (ILP) solver, we combine the LLM’s qualitative strengths in service-type clustering with the ILP’s quantitative rigor in constraint enforcement. This hybrid approach consistently achieved 100\% completeness and balanced slice utilization, outperforming either method individually.  

The results highlight a complementary relationship: the LLM can quickly narrow the solution space and provide a high-quality initial allocation, while the ILP solver guarantees feasibility and optimality. Such an AI-and-optimization workflow can significantly reduce manual effort for network operators, delivering allocation plans that are both operationally sound and computationally efficient. Future work will extend this approach to larger, dynamic, and multi-objective slicing scenarios, further advancing AI-driven automation in beyond-5G systems.

\appendix
\section{Prompt Design}\label{app:prompt}
The following is the template of the prompt we used for Claude in the zero-shot slice assignment task. (Fields in curly braces \{...\} indicate where dynamic content, like the list of slices or requests, was inserted for a specific test instance.)

\begin{tcolorbox}[colback=gray!10, colframe=gray!80, title=Prompt to Language Model, sharp corners, fonttitle=\bfseries]
\small
\textbf{System message:} You are a network slicing assistant.

\textbf{User prompt:} Network slicing is the process of partitioning a physical mobile network into multiple virtual slices to meet diverse service requirements. For example, eMBB slices provide high bandwidth for video streaming, URLLC slices offer ultra-low latency for mission-critical services, and mMTC slices support massive IoT connectivity. Telecom operators typically perform slice resource allocation manually.

Assign each of the following user service requests to one of the available network slices based on these constraints:
1. The total resource demand of all requests in a slice must not exceed that slice’s capacity.
2. No new slices should be added (use only the provided slices).
3. A request’s latency requirement must be satisfied by the slice’s latency (assign each request only to a slice with equal or lower latency than it requires).
4. All requests must be assigned to a slice (no request left unassigned).

The output should be a list of assignments in CSV format (as a data frame) enclosed in triple backticks, using the "@" symbol as a delimiter. Each line should be:

slice\_id @ request\_id @ allocated\_units

(representing that a given request is allocated to a particular slice along with the resource units it will consume).

Example output format (for illustration, assuming slice names and request IDs):

\begin{verbatim}
SliceA@Request1@5
SliceB@Request2@10
SliceA@Request3@8
...\end{verbatim}

Below are the available slices and their capacities, followed by the list of user requests with their demands and latency requirements:

Slice Capacities:
\{list of slices with capacity and latency info\}

User Requests:
\{list of requests with demand and latency requirement\}

Ensure that your response strictly follows the format and constraints above, and nothing else.
\end{tcolorbox}

\noindent We took several measures in prompt construction to improve the LLM’s output:
\begin{itemize}
    \item We provided a clear listing of slices with capacities (and latency attributes) and a list of requests with their parameters, so the LLM had all necessary data explicitly.
    \item We shuffled the order of the user requests in the input each time to discourage the LLM from learning any particular ordering or grouping from the prompt context (preventing it from just echoing a given order).
    \item We chose the "@" symbol as a separator instead of a comma to avoid confusion in CSV parsing, since some request or slice names might contain commas.
    \item We included an example of the desired output format inside triple backticks to show the LLM exactly how to format the assignment. The final instruction explicitly told the LLM to prioritize the correct format above all else.
\end{itemize}
These prompt design choices were informed by initial trials and helped maximize the chance of getting a usable, correctly-formatted solution from the LLM.

\section{Integer Programming Formulation}\label{app:IP}
To formally capture the optimization problem for slice allocation (with a similarity-based objective), we define the following decision variables:
\begin{itemize}
    \item \textbf{Assignment variables:} $x_{i,m} \in \{0,1\}$ for each request $i$ and slice $m$, where $x_{i,m} = 1$ if request $i$ is assigned to slice $m$ (and 0 otherwise). Every request will be assigned to exactly one slice in a feasible solution, so exactly one of the $x_{i,m}$ for a given $i$ should be 1.
    \item \textbf{Pairing variables:} $z_{i,j,m} \in \{0,1\}$ for each pair of requests $i, j$ (and each slice $m$), where $z_{i,j,m} = 1$ if and only if requests $i$ and $j$ are both assigned to slice $m$. These variables link pairwise assignments and will be used to compute the objective.
\end{itemize}

\subsection*{B.1 Objective Function}
We aim to maximize the total similarity among requests that are placed in the same slice. We assume we have a similarity score $\text{sim}(i,j)$ for each pair of requests (this is determined by the LLM, as discussed in the main text). If two requests $i$ and $j$ end up in the same slice and they are similar, that should contribute to the objective. We can write the objective as:
\[
\max \sum_{m=1}^{M} \sum_{i=1}^{N} \sum_{j=1}^{N} \text{sim}(i,j) \cdot z_{i,j,m} ~,
\] 
which sums the similarity score for every pair of requests that share a slice (note that each pair $i,j$ contributes only for the slice they both occupy, and $\text{sim}(i,j)=0$ for dissimilar pairs means those pairs don't add anything).

This formulation encourages the optimization to co-locate requests that the LLM (or baseline heuristic) marked as similar on the same slice, thereby implicitly aligning with service-based slicing goals (e.g., keeping all video streaming users together on an eMBB slice).

\subsection*{B.2 Constraints}
The solution must satisfy all the hard constraints of the assignment problem:

\begin{enumerate}
    \item \textbf{Linking $z$ and $x$ variables:} The $z_{i,j,m}$ variables should be consistent with the $x$ (assignment) variables. In particular, $z_{i,j,m}$ should be 1 if and only if both $x_{i,m}$ and $x_{j,m}$ are 1 (meaning requests $i$ and $j$ are both assigned to slice $m$). We enforce this with linear constraints:
    \[
    z_{i,j,m} \le x_{i,m}, \qquad
    z_{i,j,m} \le x_{j,m}, \qquad
    z_{i,j,m} \ge x_{i,m} + x_{j,m} - 1~,
    \] 
    for all requests $i,j$ and slices $m$. These ensure that $z_{i,j,m}$ can only be 1 if both corresponding $x$'s are 1, and if both $x$'s are 1 then the third inequality forces $z_{i,j,m}=1$. (We include all unordered pairs $i,j$, $i<j$, in practice to avoid duplicates, but the formulation is conceptually the same.)
    
    \item \textbf{Each request assigned exactly once:} Every request must go to one and only one slice. Thus for each request $i$,
    \[
    \sum_{m=1}^{M} x_{i,m} = 1 \quad \forall i = 1,2,\dots,N~.
    \]
    This constraint guarantees that no request is left unassigned or assigned to multiple slices.
    
    \item \textbf{Slice capacity constraints:} The total resource demand of requests in a slice cannot exceed that slice’s capacity. Let $d_i$ be the demand of request $i$ (in the same units as slice capacity). For each slice $m$,
    \[
    \sum_{i=1}^{N} d_i \, x_{i,m} \le C_m \quad \forall m = 1,2,\dots,M~,
    \] 
    where $C_m$ is the capacity of slice $m$. This ensures no slice is over-allocated. The sum of demands for all requests assigned to slice $m$ must stay within $C_m$.
    
    \item \textbf{Latency constraints:} A request can only be assigned to a slice that meets its latency requirement. We enforce this by disallowing assignments that violate latency. Suppose request $i$ requires latency $\ell_i$ (maximum tolerable delay), and slice $m$ guarantees latency $\ell^{(m)}$ (e.g., slice $m$ might have a round-trip time up to 5\,ms if it’s URLLC, or 50\,ms if it’s eMBB). If slice $m$’s latency $\ell^{(m)}$ is higher (worse) than $\ell_i$, then request $i$ cannot be placed in $m$. We implement this by precomputing an \emph{allowed assignment matrix} and adding constraints:
    \[
    x_{i,m} = 0 \qquad \text{for each request–slice pair $(i,m)$ that is incompatible due to latency.}
    \] 
    In practice, this means we either do not include those $x_{i,m}$ variables in the ILP at all, or we fix them to 0. This ensures no request is placed on an unsuitable slice in terms of latency.
\end{enumerate}

Solving the above integer program yields an optimal assignment of requests to slices that maximizes the similarity-based objective while strictly satisfying capacity and latency constraints and ensuring each request is assigned exactly once. In our experiments, we provided the similarity matrix as input (from either the LLM or the baseline method) and then used the solver to obtain the results. The formulation highlights how an LLM’s output (similarity assessments) can be mathematically integrated into a rigorous optimization framework, combining data-driven insights with a hard guarantee of feasibility.

\end{document}